\documentclass[12pt,aps,nofootinbib,showpacs,preprint,preprintnumbers]{revtex4}
\usepackage{epsfig,subfigure}

\def\DDbar{D{}^0-\overline D{}^0}
\def\BsBsbar{B_s{}^0-\overline B_s{}^0}

\def\Bbar{\overline{B}}
\def\DG{\Delta \Gamma_{B_s}}
\def\DM{\Delta M_{B_s}}

\def\beq{\begin{equation}}
\def\eeq{\end{equation}}
\def\bea{\begin{eqnarray}}
\def\eea{\end{eqnarray}}

\def\dfrac#1#2{{\displaystyle {#1 \over #2}}}
\def\vv{V^*_{cb}V^{\phantom{*}}_{cs}}

\newcommand{\bld}[1]{\mbox{\boldmath$#1$}}

\begin{document}
\preprint{\vbox {
\hbox{WSU--HEP--0702}
}}

\vspace*{1cm}

\title{\boldmath Lifetime Difference in \bld{B_s} mixing: Standard Model and beyond}

\author{Andriy Badin$^1$, Fabrizio Gabbiani$^1$ and Alexey A. Petrov$^{1,2}$}

\affiliation{$^1$Department of Physics and Astronomy, Wayne State University,
Detroit, Michigan 48201\\
$^2$Michigan Center for Theoretical Physics, University of Michigan,
Ann Arbor, Michigan 48109}


\begin{abstract}
We present a calculation of $1/m^2_b$ corrections to the lifetime differences 
of $B_s$ mesons $\DG$ in the heavy-quark expansion. We find that they are small 
to significantly affect $\DG$ and present the result for lifetime difference 
including non-perturbative $1/m_b$ and $1/m_b^2$ corrections. We also analyze 
the generic $\Delta B = 1$ New Physics contributions to $\DG$ and 
provide several examples.

\end{abstract}

\vspace*{1in}

\maketitle

\section{Introduction}

Meson-antimeson mixing serves as an indispensable way of placing constraints on 
various models of New Physics (NP). This is usually ascribed to the fact that 
this process only occurs at the one-loop level in the Standard Model (SM) of electroweak 
interactions. This makes it sensitive to the effects of possible NP particles in 
the loops or even to new tree-level interactions that can possibly contribute to the 
flavor-changing $\Delta Q=2$ interactions. These interactions induce non-diagonal 
terms in the meson-antimeson mass matrix that describes the dynamics of those states.
Diagonalizing this mass matrix gives two mass eigenstates that are superpositions of
flavor eigenstates. In the $B_s$ system mass eigenstates, denoted as ``heavy'' 
$|B_H\rangle$ and ``light'' $|B_L\rangle$,
\bea
|B_H\rangle &=& p |B_s\rangle + q |\overline{B}_s\rangle,
\nonumber \\
|B_L\rangle &=& p |B_s\rangle - q |\overline{B}_s\rangle,
\eea
were predicted to have a rather significant mass and width differences,
\beq
\Delta M_ {B_s}= M_H - M_L, \qquad \Delta \Gamma_{B_s} = \Gamma_L - \Gamma_H,
\eeq
where $M_{H,L}$ and $\Gamma_{H,L}$ denote mass and lifetime differences of
mass eigenstates. Since in the Standard Model the mass difference is dominated by 
the top quark contributions, it is computable with great accuracy. Thus one might 
expect that possible NP contributions can be easily isolated. Unfortunately, a recent 
observation of mass difference of 
mass eigenstates in $B_s$ mixing by CDF~\cite{Abulencia:2006ze} and D0~\cite{Abazov:2006dm},
\begin{eqnarray}
&& \Delta M_{B_s} = 17.77 \pm 0.10 \pm 0.07 \mbox{~ps}^{-1} \mbox{~~(CDF)},
\nonumber \\
&& 17 \mbox{~ps}^{-1} <  \Delta M_{B_s} < 21 \mbox{~ps}^{-1}  \mbox{~~~~~(D0)},
\end{eqnarray}
put the hopes of spectacular NP effects in $B_s$ system rest. In fact,
analyses of mixing in the strange, charm and beauty quark systems all yielded positive 
signals, yet all of those signals seem to be explained quite well by the SM interactions.
Yet, some contribution from New Physics particles is still possible, so even the 
energy scales above those directly accessible at the Tevatron or LHC can be probed with 
$B_s$ mixing, provided that QCD sum rule~\cite{Korner:2003zk} or lattice 
QCD~\cite{Dalgic:2006gp,Becirevic:2001xt,Gimenez:2000jj,Aoki:2003xb} 
calculations supply the relevant hadronic parameters with sufficient accuracy.

In addition to the mass difference $\DM$, a number of experimental collaborations
reported the observation of a lifetime difference $\DG$ in the $B_s$ system.
Combining recent result from D0~\cite{Abazov:2007tx} with earlier measurements 
from CDF~\cite{Acosta:2004gt} and ALEPH~\cite{Barate:2000kd}, Particle Data Group (PDG)
quotes~\cite{PDG}
\begin{equation}
\DG = 0.16^{+0.10}_{-0.13} \mbox{~ps}^{-1}, \qquad 
\frac{\DG}{\Gamma_{B_s}} = 0.121^{+0.083}_{-0.090},
\end{equation}
while Heavy Flavor Averaging Group (HFAG)~\cite{Barberio:2007cr} gives
\begin{equation}
\DG = 0.071^{+0.053}_{-0.057} \mbox{~ps}^{-1}, \qquad 
\frac{\DG}{\Gamma_{B_s}} = 0.104^{+0.076}_{-0.084}.
\end{equation}
Differently from the mass difference $\DM$, the lifetime difference
$\DG$ is definitely dominated by the SM contributions, as it is
generated by the on-shell 
intermediate states~\cite{Beneke:1996gn, Beneke:1998sy,Ciuchini:2003ww,Lenz:2006hd}. 
While this might appear to make it less exciting for indirect searches for New
Physics, besides ``merely'' providing yet another test for heavy quark expansion, 
it is nonetheless a useful quantity for a combined analysis of possible NP 
contributions to $\BsBsbar$
mixing~\cite{Grossman:2006ce,Ligeti:2006pm,Grossman:1996er,Dunietz:2000cr}.

It has been argued~\cite{Grossman:1996er} that CP-violating NP contributions 
to $\Delta B=2$ amplitudes can only reduce the experimentally-observed lifetime 
difference compared to its SM value, therefore it is important to have an accurate 
theoretical evaluation of $\DG$ in the SM. It is also important to note that $\Delta B = 1$ 
NP contributions can affect $\DG$, but do not have to follow
the same pattern. Indeed, the level at which $\Delta B = 1$ NP can affect $\DG$ 
depends both on the particular extension of the SM, as well as on the projected accuracy 
of lattice calculations of hadronic parameters which drives the uncertainties on the 
theoretical prediction of $\DG$. 
So it is advantageous to evaluate the effect of NP
contributions.

This paper is organized as follows. We set up the relevant formalism and argue for
the need to compute $1/m^2_b$ corrections to leading and next-to-leading effects in 
Sect.~\ref{formalism}. In Sect.~\ref{AllResults} we discuss the impact of $1/m^2_b$ 
corrections to the lifetime differences of $B_s$ mesons and assess the convergence of 
the $1/m_b$ expansion. We also present the complete SM results for $\DG$ including 
$1/m_b^2$ corrections. We then discuss the possible effects from $\Delta B = 1$ New Physics
contributions in Sect.~\ref{NewPhysics}. Finally, we present our
conclusions in Sect.~\ref{Conclusions}.

\section{Formalism}\label{formalism}

In the limit of exact CP conservation the mass eigenstates of the
$B_s^0$--$\overline{B}^0_s$ system are
$|B_{H/L}\rangle=(|B_s\rangle\pm |\Bbar_s\rangle)/\sqrt{2}$, with the
convention $CP|B_s\rangle=-|\Bbar_s\rangle$. The width difference
between mass eigenstates is then given by~\cite{Beneke:1996gn}
\beq
\label{dgdef}
\Delta \Gamma_{B_s}\equiv\Gamma_L-\Gamma_H =
-2\,\Gamma_{12}=-2\,\Gamma_{21},
\eeq
where $\Gamma_{ij}$ are the elements of the decay-width matrix,
$i,j=1,2$
($|1\rangle=|B_s\rangle$, $|2\rangle=|\Bbar_s\rangle$).

We use the optical theorem to relate the
off-diagonal elements of the decay-width matrix $\Gamma$ entering the neutral
$B$-meson oscillations to the imaginary part of the forward matrix
element of the transition operator ${\cal T}$:
\beq\label{rate}\label{tdef}
\Gamma_{21}(B_s)=\frac{1}{2 M_{B_s}} \langle \overline{B}_s |{\cal T} | B_s \rangle,\quad
{\cal T} = {\mbox{Im}}~ i \int d^4 x T \left\{
H_{\mbox{\scriptsize eff}}(x) H_{\mbox{\scriptsize eff}}(0) \right \}.
\eeq
Here ${\cal H}_{eff}$ is the low energy effective weak
Hamiltonian mediating bottom-quark decays. The component that is
relevant for $\Gamma_{21}$ reads explicitly
\begin{equation}\label{hpeng}
{\cal H}_{eff}=\frac{G_F}{\sqrt{2}}\vv
\left(\,\sum^6_{r=1} C_r Q_r + C_8 Q_8\right),
\end{equation}
defining the operators
\begin{equation}\label{q1q2}
Q_1= (\bar b_ic_j)_{V-A}(\bar c_js_i)_{V-A},\qquad
Q_2= (\bar b_ic_i)_{V-A}(\bar c_js_j)_{V-A},
\end{equation}
\begin{equation}\label{q3q4}
Q_3= (\bar b_is_i)_{V-A}(\bar q_jq_j)_{V-A},\qquad
Q_4= (\bar b_is_j)_{V-A}(\bar q_jq_i)_{V-A},
\end{equation}
\begin{equation}\label{q5q6}
Q_5= (\bar b_is_i)_{V-A}(\bar q_jq_j)_{V+A},\qquad
Q_6= (\bar b_is_j)_{V-A}(\bar q_jq_i)_{V+A},
\end{equation}
\begin{equation}\label{q8}
Q_8= \frac{g}{8\pi^2}m_b\,
\bar b_i\sigma^{\mu\nu}(1-\gamma_5)T^a_{ij} s_j\, G^a_{\mu\nu}.
\end{equation}
Here $i,j$ are color indices and a summation over
$q=u$, $d$, $s$, $c$, $b$ is implied.
$V\pm A$ refers to $\gamma^\mu(1\pm\gamma_5)$ and $S-P$ (which
we need below) to $(1-\gamma_5)$.
$C_1,\ldots, C_6$ are the corresponding Wilson coefficient
functions at the renormalization scale $\mu$, which are known at
next-to-leading order. We have also included the chromomagnetic
operator $Q_8$, contributing to ${\cal T}$ at ${\cal O}(\alpha_s)$.
Note that for a negative $C_8$, as conventionally used in the literature,
the Feynman rule for the quark-gluon vertex
is $-ig\gamma_\mu T^a$.
A detailed review and explicit expressions may be found in~\cite{Buchalla:1995vs}.
Cabibbo-suppressed channels have been neglected in Eq.~(\ref{hpeng}).

In the heavy-quark limit, the energy release supplied by the b-quark is large, 
so the correlator in Eq.~(\ref{rate}) is dominated by short-distance physics~\cite{Shifman:1984wx}.
An Operator Product Expansion (OPE) can be constructed for Eq.~(\ref{rate}),
which results in its expansion as
a series of matrix elements of local operators of increasing dimension
suppressed by powers of $1/m_b$:
\beq\label{expan}
\Gamma_{21}(B_s)= \frac{1}{2 M_{B_s}} \sum_k \langle B_s |{\cal T}_k | B_s \rangle
=\sum_{k} \frac{C_k(\mu)}{m_b^{k}}
\langle B_s |{\cal O}_k^{\Delta B=2}(\mu) | B_s \rangle.
\eeq
In other words, the calculation of $\Gamma_{21}(B_s)$ is equivalent to computing
the matching coefficients of the effective $\Delta B=2$ Lagrangian
with subsequent computation of its matrix elements. Eventually the
scale dependence of the Wilson coefficients in Eq.~(\ref{expan}) is
bound to match the scale dependence of the computed matrix elements.

Expanding the operator product (\ref{tdef}) for small $x\sim 1/m_b$,
the transition operator ${\cal T}$ can be written to leading order
in the $1/m_b$ expansion as~\cite{Beneke:1996gn,Beneke:1998sy}
\beq\label{tfq}
{\cal T}=-\frac{G^2_F m^2_b}{12\pi}(\vv)^2
\, \left[ F(z) Q(\mu_2)+ F_S(z) Q_S(\mu_2) \right],
\eeq
which results in~\cite{Ciuchini:2003ww}
\bea\label{tres}
&\Gamma&_{21}(B_s) = -\frac{G^2_F m^2_b}{12\pi (2 M_{B_s})}(\vv)^2
\sqrt{1-4z}\times  \nonumber\\
&\times&\left\{\left[(1-z)\,\left(2\, C_1 C_2+N_c C^2_2\right)+(1-4z)
C^2_1/2 \right] \langle Q \rangle +
(1+2z)\left(2\, C_1 C_2+N_c C^2_2-C^2_1\right) \langle Q_S \rangle \right\},
\eea
with $z=m_c^2/m_b^2$ and the basis of $\Delta B=2$ operators\footnote{It was 
recently argued that better-converging results can be obtained in a 
modified basis~\cite{Lenz:2006hd}.}
\beq\label{qqs}
Q = (\bar b_is_i)_{V-A}(\bar b_js_j)_{V-A},\qquad
Q_S= (\bar b_is_i)_{S-P}(\bar b_js_j)_{S-P} ~.
\eeq
In writing Eq.~(\ref{tfq}) we have used the Fierz identities and the equations
of motion to eliminate the color re-arranged operators
\beq\label{qqt}
\tilde Q=(\bar b_is_j)_{V-A}(\bar b_js_i)_{V-A},\qquad
\tilde Q_S= (\bar b_is_j)_{S-P}(\bar b_js_i)_{S-P}~,
\eeq
always working to leading order in $1/m_b$. Note that
$\langle ... \rangle$ denote matrix elements of the above operators
taken between $B_s$ and ${\overline B}_s$ states.
The Wilson coefficients $F$ and $F_S$ can be extracted
by computing the matrix elements between quark states of
${\cal T}$ in Eq.~(\ref{tdef}). 

The coefficients in the transition operator (\ref{tfq})
at next-to-leading order, still neglecting the penguin sector,
can be written as~\cite{Beneke:1998sy}:
\beq\label{fz}
F(z)=F_{11}(z) C^2_2(\mu_1)+ F_{12}(z) C_1(\mu_1) C_2(\mu_1)+
     F_{22}(z) C^2_1(\mu_1),
\end{equation}
\begin{equation}\label{fij}
F_{ij}(z)=F^{(0)}_{ij}(z)
+\frac{\alpha_s(\mu_1)}{4\pi}F^{(1)}_{ij}(z),
\eeq
and similarly for $F_S(z)$.
The leading order functions $F^{(0)}_{ij}$, $F^{(0)}_{S,ij}$
read explicitly
\begin{equation}\label{f011}
F^{(0)}_{11}(z)=3\sqrt{1-4z}\, (1-z),\qquad
F^{(0)}_{S,11}(z)=3\sqrt{1-4z}\, (1+2z),
\end{equation}
\begin{equation}\label{f012}
F^{(0)}_{12}(z)=2\sqrt{1-4z}\, (1-z),\qquad
F^{(0)}_{S,12}(z)=2\sqrt{1-4z}\, (1+2z),
\end{equation}
\begin{equation}\label{f022}
\,F^{(0)}_{22}(z)=\frac{1}{2}(1-4z)^{3/2},\qquad\qquad
F^{(0)}_{S,22}(z)=-\sqrt{1-4z}\, (1+2z).
\end{equation}
The next-to-leading order (NLO) expressions
$F^{(1)}_{ij}$, $F^{(1)}_{S,ij}$ are given in Ref.~\cite{Beneke:1998sy}.

The penguin correction to Eq.~(\ref{tfq}),
\begin{equation}\label{tpcq}
{\cal T}_p=-\frac{G^2_F m^2_b}{12\pi}(\vv)^2
\, \left[ P(z) Q+ P_{S}(z) Q_S \right],
\end{equation}
is also shown in Ref.~\cite{Beneke:1998sy}.

\section{Subleading $1/m_b^n$ corrections}\label{AllResults}

Here we present the higher order corrections
to $\Gamma_{21}(B_s)$ in Eq.~(\ref{tres}) in the heavy-quark expansion,
denoted below as $\delta_{1/m}$ and $\delta_{1/m^2}$:
\beq\label{corr}
\Gamma_{21}(B_s) = -\frac{G^2_F m^2_b}{12\pi (2 M_{B_s})}(\vv)^2
\,\left\{\left[F(z)+P(z)\right] \langle Q \rangle + 
\left[F_S(z)+P_S(z)\right] \langle Q_S \rangle
+\delta_{1/m} + \delta_{1/m^2}\right\}\,.
\eeq
The matrix elements for $Q$ and $Q_S$ are known to
be~\cite{Beneke:1996gn,Beneke:1998sy,Ciuchini:2003ww}
\bea\label{meqs}
\langle Q \rangle &\equiv& \langle \overline B_s\vert Q\vert B_s\rangle 
= f^2_{B_s}M^2_{B_s}2\left(1+\frac{1}{N_c}\right)B,
\nonumber \\[0.1cm]
\langle Q_S \rangle &\equiv& \langle \overline B_s\vert Q_S \vert B_s\rangle 
= -f^2_{B_s}M^2_{B_s}
\frac{M^2_{B_s}}{(m_b+m_s)^2}\left(2-\frac{1}{N_c}\right)B_S,
\\
\delta_{1/m} &=& \langle \overline B_s\vert {\cal T}_{1/m} \vert B_s\rangle, 
\qquad \mbox{and} \qquad 
\delta_{1/m^2} = \langle \overline B_s\vert {\cal T}_{1/m^2} \vert B_s\rangle,
\nonumber
\eea
\noindent where $M_{B_s}$ and $f_{B_s}$ are the mass and decay constant
of the $B_s$ meson and $N_c$ is the number of colors.
The parameters $B$ and $B_S$ are defined such that
$B=B_S=1$ corresponds to the factorization (or `vacuum insertion') approach, 
which can provide a first estimate.

\subsection{\mbox{\boldmath $1/m_b$} corrections}

\begin{figure}[tb]
\centerline{\epsfxsize=9cm\epsffile{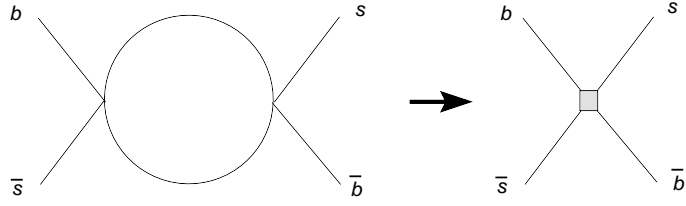}}
\centerline{\parbox{14cm}{\caption{\label{fig:KineticOper}
Calculation of kinetic $1/m_b$ and $1/m_b^2$ corrections. The operators 
of Eqs.~(\ref{OurCorrection})
and (\ref{OurCorrection1}) are obtained by expanding the diagrams in powers of
light quark momentum.}}}
\end{figure}

The $1/m_b$ corrections are computed, as in 
Ref.~\cite{Beneke:1996gn,Ciuchini:2003ww,Shifman:1984wx,Gabbiani:2004tp},
by expanding the forward scattering amplitude of Eq.~(\ref{rate}) in the light-quark momentum
and matching the result onto the operators containing derivative insertions 
(see Fig.~\ref{fig:KineticOper}). The $\delta_{1/m}$ contributions can be written in
the following form:
\bea\label{OurCorrection}
{\cal T}_{1/m} &=& \sqrt{1-4z}\left\{
(1 + 2 z)\left[C^2_1\, \left(R_2 + 2 R_4\right) -2\, (2\, C_1 C_2+N_c C_2^2)
\,\left(R_1+R_2\right)\right]\phantom{\frac{z^2}{4}}\right.\nonumber \\
&-&
\left.\frac{12 z^2}{1 - 4z}\left[(2\, C_1 C_2+N_c C^2)\,\left(R_2+2 R_3\right) + 2\, C^2_1
\,R_3\right]\right\},
\eea
where the operators $R_i$ are defined as
\bea
R_1 &=& \dfrac{m_s}{m_b}\bar b_i \gamma^{\mu} (1-\gamma_5) s_i
~\bar b_j \gamma_{\mu} (1+\gamma_5) s_j\,, \qquad\ \
R_2 = \dfrac{1}{m^2_b}\bar b_i {\overleftarrow D}_{\!\rho}
\gamma^\mu (1-\gamma_5)\overrightarrow{D}^\rho s_i
~\bar b_j\gamma_\mu(1-\gamma_5)s_j\,,\nonumber \\
R_3 &=& \dfrac{1}{m^2_b}\bar b_i{\overleftarrow D}_{\!\rho}
(1-\gamma_5)\overrightarrow{D}^\rho s_i~\bar b_j(1-\gamma_5)s_j\,,\qquad
R_4 = \dfrac{1}{m_b}\bar b_i(1-\gamma_5)i\overrightarrow{D}_\mu s_i
~\bar b_j\gamma^\mu(1-\gamma_5)s_j\,.
\label{rrt1}
\eea
Their matrix elements read~\cite{Beneke:1996gn,Ciuchini:2003ww}:
\bea\label{KinME1}
\langle  \overline  B_s | R_1 | B_s \rangle  &=&
\left(2+\frac{1}{N_c}\right)\,\dfrac{m_s}{m_b}\, f_{B_s}^2 M_{B_s}^2\,B^{s}_1,\quad\quad
\langle  \overline  B_s | R_2 | B_s \rangle   =
\left(-1+\frac{1}{N_c}\right)\,f_{B_s}^2 M_{B_s}^2
\left( \dfrac{M_{B_s}^2}{m_b^2} - 1 \right)\,B^{s}_2 
\nonumber,\\
\langle  \overline  B_s | R_3 | B_s \rangle  &=&
\left(1+\frac{1}{2N_c}\right)\,f_{B_s}^2 M_{B_s}^2
\left( \dfrac{M_{B_s}^2}{m_b^2} - 1
\right)\,B^{s}_3,\quad
\langle  \overline  B_s | R_4 | B_s \rangle =
-f_{B_s}^2 M_{B_s}^2
\left( \dfrac{M_{B_s}^2}{m_b^2} - 1 \right)\,B^{s}_4\,.
\eea
Some of those parameters have been computed in lattice 
QCD~\cite{Dalgic:2006gp,Becirevic:2001xt,Gimenez:2000jj,Aoki:2003xb}.\footnote{For estimates 
of these matrix elements based on QCD sum rules, see Ref.~\cite{Korner:2003zk}.}
In this paper we use the results of Ref.~\cite{Dalgic:2006gp}.

The color-rearranged operators $\widetilde{R}_i$ that follow from the expressions for $R_i$ by
interchanging color indexes of $b_i$ and $s_j$ Dirac spinors have been
eliminated using Fierz identities and the equations
of motion as in Eq.~(\ref{qqs}).
Note that the above result contains {\it full} QCD $b$-fields, thus there
is no immediate power counting available for these operators. The power
counting becomes manifest at the level of the matrix elements.

\subsection{\mbox{\boldmath $1/m^2_b$} corrections}

It was shown in Refs.~\cite{Beneke:1996gn,Ciuchini:2003ww} that $1/m_b$-corrections
are quite large, so it is important to assess the convergence of $1/m_b$-expansion 
in the calculation of the $B_s$ lifetime difference. In order to do so, we compute 
a set of $\delta_{1/m_b^2}$ corrections to leading order. As
expected, at this order more operators will contribute.  We will
parametrize the $1/m_b^2$ corrections similarly to our parametrization
of $1/m_b$ effects above and use the factorization approximation
to assess their contributions to the $B_s$ lifetime difference.
\begin{figure}[tb]
\centerline{\epsfxsize=9cm\epsffile{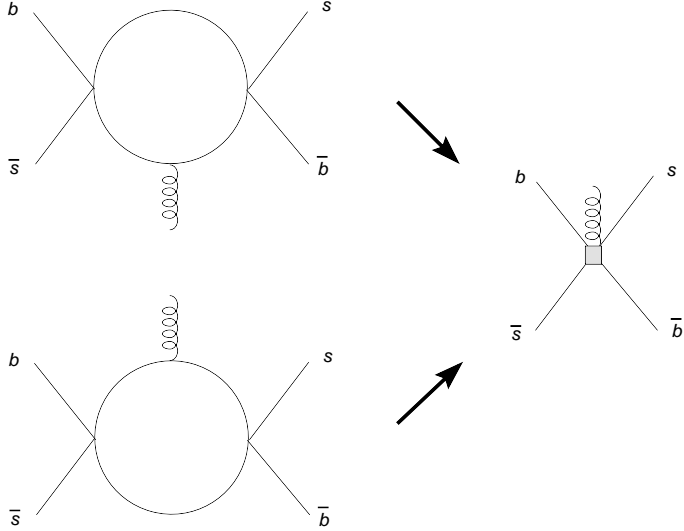}}
\centerline{\parbox{14cm}{\caption{\label{fig:GluonicOper}
$1/m_b^2$-corrections from gluonic operators.}}}
\end{figure}

Two classes of corrections arise at this order. One class involves kinetic
corrections which can be computed in a way analogous to the previous case by expanding the
forward scattering amplitudes in the powers of the light-quark momentum.
A second class involves corrections arising from the interaction with background gluon
fields. The complete set of corrections is the sum of those, 
\beq
{\cal T}_{1/m^2} = {\cal T}_{1/m^2}^{kin} + {\cal T}_{1/m^2}^{G}.
\eeq
Let us consider those classes of corrections in turn. The kinetic corrections can be written as
\bea\label{OurCorrection1}
{\cal T}_{1/m^2}^{kin} &=& \sqrt{1-4z} \Bigl[
\dfrac{24 z^2}{(1-4 z)^2}(3-10z)\left[
C_1^2 W_3+(2\, C_1 C_2 +N_c C_2^2) (W_3+W_2/2)\right]
\nonumber \\
&+&\dfrac{12 z^2}{1-4 z}\dfrac{m_s^2}{m_b^2}\left[
C_1^2 Q_S-(2\, C_1 C_2 +N_c C_2^2) (Q_S+Q/2)\right]
\nonumber \\
&+&\dfrac{24 z^2}{1-4 z}\left[
2 C_1^2 W_4-2\, (2\, C_1 C_2 +N_c C_2^2) (W_1+W_2/2)\right]
\nonumber \\
&-&(1-2 z)\dfrac{m_s^2}{m_b^2}(C_1^2+2\, C_1 C_2 +N_c C_2^2)Q_R \Bigr].
\eea
We again retain the dependence on quark masses in the above expression, including the
terms proportional to $m_s$.
The operators in Eq.~(\ref{OurCorrection1}) are defined as
\bea
Q_R &=& (\bar b_is_i)_{S+P}(\bar b_js_j)_{S+P}\,,\nonumber \\
W_1 &=& \dfrac{m_s}{m_b}
\bar b_i\overleftarrow{D}^{\alpha}(1-\gamma_5)\overrightarrow{D}_{\alpha}s_i~\bar b_j(1+\gamma_5)s_j\,,
\nonumber \\
W_2 &=& \dfrac{1}{m_b^4}
\bar b_i\overleftarrow{D}^{\alpha}\overleftarrow{D}^{\beta}
\gamma^{\mu}(1-\gamma_5)\overrightarrow{D}_{\alpha}\overrightarrow{D}_{\beta}s_i
~\bar b_j\gamma_{\mu}(1-\gamma_5)s_j\,,  
\nonumber \\
W_3 &=& \dfrac{1}{m_b^4}
\bar b_i\overleftarrow{D}^{\alpha}\overleftarrow{D}^{\beta}
(1-\gamma_5)\overrightarrow{D}_{\alpha}\overrightarrow{D}_{\beta}s_i
~\bar b_j (1-\gamma_5) s_j\,, \nonumber \\
W_4 &=& \dfrac{1}{m_b^4}
\bar b_i\overleftarrow{D}^{\alpha}(1-\gamma_5)i \overrightarrow{D}_{\mu}\overrightarrow{D}_{\alpha}s_i
~\bar b_j\gamma^{\mu}(1-\gamma_5)s_j\,,
\eea
where, as before, we have eliminated the color-rearranged operators
$\widetilde{W}_i$ in favor of the operators $W_i$.
The parametrization of the matrix elements of the above
operators is given below,
\bea\label{KinME2}
\langle \overline  B_s |Q_R| B_s \rangle &=& -\left(2-\frac{1}{N_c}\right) 
f^2_{B_s}M^2_{B_s} \frac{M^2_{B_s}}{(m_b+m_s)^2} \alpha_1 \,, 
\nonumber \\
\langle \overline  B_s |W_1| B_s \rangle &=& \dfrac{m_s}{m_b} 
\left(1+\frac{1}{2 N_c}\right)\,f_{B_s}^2 M_{B_s}^2
\left( \dfrac{M_{B_s}^2}{m_b^2} - 1 \right)\alpha_2\,, 
\nonumber \\
\langle \overline  B_s |W_2| B_s \rangle &=& \frac{1}{2}\left(-1+\frac{1}{N_c}\right)\,f_{B_s}^2 M_{B_s}^2
\left( \dfrac{M_{B_s}^2}{m_b^2} - 1 \right)^2 \alpha_3\,, 
\nonumber \\
\langle \overline  B_s |W_3| B_s \rangle &=& \frac{1}{2} \left(1+\frac{1}{2 N_c}\right)\,f_{B_s}^2 M_{B_s}^2
\left( \dfrac{M_{B_s}^2}{m_b^2} - 1 \right)^2 \alpha_4 \,, \nonumber \\
\langle \overline  B_s |W_4| B_s \rangle &=& -\frac{1}{2}\,f_{B_s}^2 M_{B_s}^2
\left( \dfrac{M_{B_s}^2}{m_b^2} - 1 \right)^2 \alpha_5 \,.
\eea
Note that in factorization approximation all the bag parameters $\alpha_i$ should be 
set to 1. In addition to the set of kinetic corrections considered above, the effects of the
interactions of the intermediate quarks with background gluon fields should also
be included at this order. 
The contribution of those operators can be computed
from the diagram of Fig.~\ref{fig:GluonicOper}, resulting in
\bea\label{Tglue}
{\cal T}_{1/m^2}^{G} =
&-& \dfrac{G_F^2(\vv)^2}{4\pi\sqrt{1-4z}}
\left\{C_1^2\left[(1-4z) P_1-(1-4z) P_2+4z P_3-4z P_4 \right]\right. \nonumber \\
&+& \left. 4\; C_1 C_2 z\left[P_5+P_6-P_7-P_{8}\right]
\phantom{C^2_1}\hskip -15pt \right\}.
\eea
The local four-quark operators in the above formulas
are shown in Eq.~(\ref{poper}):
\bea \label{poper}
P_1 &=& \bar b_i\gamma^{\mu}(1-\gamma_5)s^{\phantom{l}}_i~
\bar b^{\phantom{l}}_k\gamma^{\nu}(1-\gamma_5)t_{kl}^a\widetilde{G}_{\mu\nu}^a s^{\phantom{l}}_l\,,
\nonumber \\
P_2 &=& \bar b_k\gamma^{\mu}(1-\gamma_5)t_{kl}^a\widetilde{G}_{\mu\nu}^a s_l~
\bar b^{\phantom{l}}_i\gamma^{\nu}(1-\gamma_5)s^{\phantom{l}}_i\,,  \nonumber\\
P_3 &=& \dfrac{1}{m_b^2}\bar b_i\overleftarrow{D}^{\mu}\overleftarrow{D}^{\alpha}\gamma^{\alpha}(1-\gamma_5)
s^{\phantom{l}}_i~
\bar b^{\phantom{l}}_k\gamma_{\nu}(1-\gamma_5)t_{kl}^a\widetilde{G}_{\mu\nu}^a s^{\phantom{l}}_l\,,
\nonumber\\
P_4 &=& \dfrac{1}{m_b^2}\bar b_k\overleftarrow{D}^{\nu}\overleftarrow{D}^{\alpha}\gamma^{\mu}(1-\gamma_5)
t_{kl}^a\widetilde{G}_{\mu\nu}^a s^{\phantom{l}}_l
~\bar b^{\phantom{l}}_i\gamma_{\alpha}(1-\gamma_5)s^{\phantom{l}}_i\,, \nonumber \\
P_5 &=& \dfrac{1}{m_b^2}\bar b_k\overleftarrow{D}^{\nu}\overleftarrow{D}^{\alpha}\gamma^{\mu}(1-\gamma_5)
s^{\phantom{l}}_i~t_{kl}^a\widetilde{G}_{\mu\nu}^a
~\bar b^{\phantom{l}}_i\gamma_{\alpha}(1-\gamma_5)s^{\phantom{l}}_l\,, \nonumber \\
P_6 &=& \dfrac{1}{m_b^2}\bar b_i\overleftarrow{D}^{\nu}\overleftarrow{D}^{\alpha}\gamma^{\mu}(1-\gamma_5)
s^{\phantom{l}}_k~t_{kl}^a\widetilde{G}_{\mu\nu}^a
~\bar b^{\phantom{l}}_l\gamma_{\alpha}(1-\gamma_5) s^{\phantom{l}}_i\,,
\nonumber \\
P_7 &=& \dfrac{1}{m_b^2}\bar b_k\overleftarrow{D}^{\mu}\overleftarrow{D}^{\alpha}\gamma^{\alpha}(1-\gamma_5)
s^{\phantom{l}}_i~t_{kl}^a\widetilde{G}_{\mu\nu}^a
~\bar b^{\phantom{l}}_i\gamma_{\nu}(1-\gamma_5)s^{\phantom{l}}_l\,,
\nonumber \\
P_8 &=& \dfrac{1}{m_b^2}\bar b_i\overleftarrow{D}^{\mu}\overleftarrow{D}^{\alpha}\gamma^{\alpha}(1-\gamma_5)
s^{\phantom{l}}_k~t_{kl}^a\widetilde{G}_{\mu\nu}^a
~\bar b^{\phantom{l}}_l\gamma_{\nu}(1-\gamma_5) s^{\phantom{l}}_i.
\eea
Analogously to the previous section, and following
Ref.~\cite{Gabbiani:2004tp}, we parametrize the matrix elements
in Eq.~(\ref{poper}) as
\beq\label{GlueME}
\langle \overline B_s | P_i | B_s \rangle = \dfrac{1}{4} f^2_{B_{s}} M^2_{B_{s}}
\left(\dfrac{M_{B_s}^2}{m^2_b}-1\right)^2 \beta_i.
\eeq
We set $\beta_i$ = 1$~GeV^2$ to obtain a numerical estimate
of this effect. It is clear that no precise prediction is possible with so many operators
contributing to the lifetime difference. This, of course, is expected, as
the number of contributing operators always increases significantly with each
order in OPE. We can nonetheless evaluate the contribution of both $1/m_b$ and $1/m_b^2$ by
randomly varying the parameters describing the matrix elements by $\pm 30\%$ around 
their ``factorized'' values. This way we obtain the interval of predictions of $\DG$ and 
estimate the uncertainty of our result.

\subsection{Discussion}

Now we discuss the phenomenological implications of the results presented in
the previous sections. As usual in OPE-based calculations next-order corrections
bring new unknown coefficients. In our numerical results we assume the value of
the $b$-quark pole mass to be $m_b=4.8\pm 0.2$~GeV and $f_{B_s}$ = 230 $\pm$ 25$~$MeV.
It might be advantageous to see what effects higher-order $1/m_b^2$ corrections have
on the value of $\DG$. In order to see that we fix all perturbative parameters at
the middle of their allowed ranges and show the dependence of $\DG$ on non-perturbative 
parameters defined in Eqs.~(\ref{KinME1}), (\ref{KinME2}), and (\ref{GlueME}):
\bea\label{parameters}
\DG \ &=& \ \Bigl[0.0005 B + 0.1732 B_s + 0.0024 B_1 - 0.0237 B_2 - 0.0024 B_3 - 0.0436 B_4
\nonumber \\ 
&+& \ 2 \times 10^{-5} \alpha_1 + 4 \times 10^{-5} \alpha_2 + 4 \times 10^{-5} \alpha_3 
+ 0.0009 \alpha_4 - 0.0007 \alpha_5 
\\
&+& \ 0.0002 \beta_1 - 0.0002 \beta_2 + 6 \times 10^{-5} \beta_3 -
6 \times 10^{-5} \beta_4 - 1 \times 10^{-5} \beta_5 
\nonumber \\
&-& \ 1 \times 10^{-5} \beta_6 + 1 \times 10^{-5} \beta_7 + 1 \times 10^{-5} \beta_8 
\Bigr]
\quad (\mbox{ps}^{-1}).
\nonumber
\eea
As one can see, $1/m_b^2$ corrections provide rather minor overall impact on the calculation
of $\DG$. In particular, contributions of gluonic operators are essentially negligible.

To obtain the complete Standard Model estimate of $\DG$, we fix the perturbative scale in our 
calculations to $\mu=m_b$ and vary the values of parameters of the matrix elements. 
Following~\cite{Gabbiani:2004tp} we adopt the statistical approach for presenting our 
results and generate 100000-point probability distributions of the lifetime, obtained by 
randomly varying our parameters within a $\pm 30\%$ interval around their 
``factorization'' values. The decay constant $f_{B_s}$ and the b-quark pole 
mass $m_b$ are taken to vary within a $1\sigma$ interval as indicated above.
The results are presented in Fig.~\ref{fig:SM}. This figure represents 
the main result of this paper~\cite{ICHEP2006}.
\begin{figure}
\centerline{\psfig{file=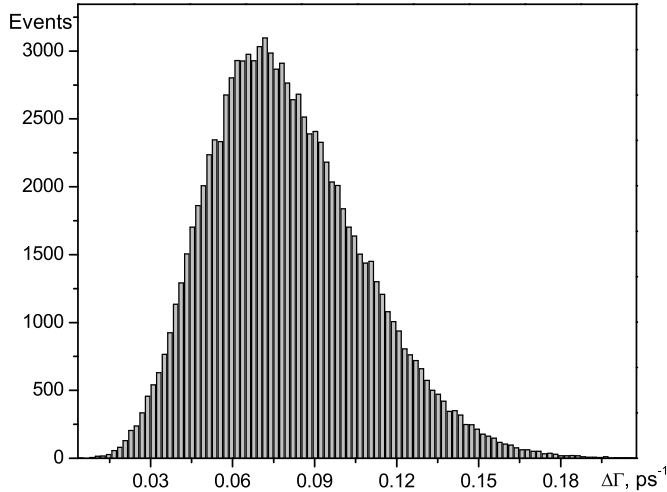,width=10cm}}
\centerline{\parbox{10cm}{\caption{\label{fig:SM} Histogram showing the random
distribution around the central values of various parameters of 
Eqs.~(\ref{corr}, \ref{OurCorrection}, \ref{OurCorrection1}, \ref{Tglue}) contributing to 
$B_s$-lifetime difference $\DG$.}}}
\end{figure}

There is no theoretically-consistent way to translate the histogram of 
Figure~\ref{fig:SM} into numerical predictions for $\DG$. As a useful estimate 
we give a numerical prediction by estimating the width of the distribution 
Fig.~\ref{fig:SM} at the middle of its height and position of the maximum of the 
curve as the most probable value. We caution that predictions obtained this way 
should be treated with care, as it is not expected that the theoretical predictions 
are distributed according to the Gaussian distribution. Nevertheless, following the 
procedure described above one obtains
\beq
\DG=0.072^{+0.034}_{-0.030} ~\mbox{ps}^{-1}, \quad \frac{\DG}{\Gamma_{B_s}} = 0.104 \pm 0.049,
\eeq
where we added the experimental error from the determination of $\Gamma_s$ and
theoretical error from our calculation of $\DG$ in quadrature.

\section{New Physics contributions to lifetime difference}\label{NewPhysics}

In the previous section we have shown that $1/m_b^2$-corrections to
the lifetime difference of the light and heavy eigenstates in the $B_s$ system
are quite small, which makes the prediction of $\DG$ quite 
reliable~\footnote{As was argued in Ref.~\cite{Lenz:2006hd}, perturbative scale dependence 
can be further reduced by switching to a different basis of leading-order operators.}. 
Additionally improving the accuracy of the lattice or QCD sum rule determinations of 
non-perturbative ``bag parameters'' in Eq.~(\ref{parameters}) would make this prediction
even more solid.

In this respect, it might be interesting to consider the effects of 
New Physics on the lifetime difference in $B_s$ system. Why would it be 
worthwhile to perform this exercise, especially since it is known that $\DG$ is dominated 
by the on-shell, real intermediate states? Wouldn't $\Delta B = 1$ New Physics amplitudes 
that can potentially affect $\DG$ already show up in the experimental studies of exclusive 
$B_s$ decays? This is indeed so. However, it might be difficult to separate New Physics
effects from the dominant (but somewhat uncertain) Standard Model contributions, as 
theoretical control over soft QCD effects is harder to achieve in the calculations of 
exclusive decays despite recent significant advances in this area~\cite{Beneke:1999br}. 

It was recently pointed out that NP contributions can dominate lifetime difference in 
$\DDbar$ system in the flavor $SU(3)$ limit~\cite{Golowich:2006gq}. In that system
this effect can be traced to the fact that the SM contribution vanishes in that limit.
While similar effect does not occur in $B_s$ mixing, good theoretical control over 
non-perturbative uncertainties in the calculation of $\DG$ makes calculations of NP 
contributions worthwhile. In $B_s$-system one can show that
\bea
\DM \ &=& \ 2 \left|M_{12}\right|, \nonumber \\
\DG \ &=& \ \frac{4 Re\left(M_{12} \Gamma^*_{12}\right)}{\DM}.
\eea
In the Standard Model the phase difference between the mixing amplitude and the dominant 
decay amplitudes is $\arg\left(-V_{cb}^* V_{cs}/V_{tb}^* V_{ts}\right)$, i.e. essentially zero.
If NP contribution has a CP-violating phase that exceeds that of the Standard Model,
one can write, denoting $2\xi = \arg\left(M_{12} \Gamma^*_{12}\right)$, 
\beq
\DG = 2 \left|\Gamma_{12}\right| \cos 2\xi.
\eeq
Since in the Standard Model $\Gamma_{12}$ is dominated by the $b \to c \bar c s$ transition,
its phase is negligible. Then, as was pointed out long time 
ago~\cite{Grossman:1996er,Dunietz:2000cr}, CP-violating contributions to $M_{12}$ must 
{\it reduce} the lifetime difference in $B_s$-system,
\beq
\DG = \DG^{SM}\cos 2\xi,
\eeq
where $2\xi$ is a CP-violating phase of $M_{12}$, which is assumed
to be dominated by some $\Delta B = 2$ New Physics.

Contrary to CP-violating $\Delta B = 2$ NP contributions to $M_{12}$, 
any $\Delta B=1$ NP amplitudes can interfere with the Standard Model ones 
both constructively and destructively, depending on the model. Since no spectacular 
NP phases have been observed in $B_s$ mixing, it
appears that $M_{12}$ is dominated by the Standard Model CP-conserving contribution.
In that case, the phase 
$\arg\left(M_{12} \Gamma^*_{12}\right) = \arg\left(\Gamma^*_{12}\right) = 2 \xi'$ is
dominated by the phase of New Physics contribution to $\Gamma^*_{12}$. In that case
\beq
\DG = \DG^{SM} + \DG^{NP} \cos 2\xi',
\eeq
where $\DG^{NP}$ is a contribution resulting form the interference of the SM and NP 
$\Delta B = 1$ operators, which can either enhance or suppress $\DG$ compared to the 
Standard Model contribution. We shall compute $\DG^{NP}$ by first employing the generic 
set of effective operators, and then specifying to particular extensions of the SM. We 
shall concentrate on CP-conserving contributions.

Using the completeness relation the NP contribution to the 
$B^0_s$-${\overline B}^0_s$ lifetime difference becomes
\begin{eqnarray}\label{gammaope}
\left. \frac{\DG}{\Gamma_{B_s}}\right|_{NP} \ &=& \
  \frac{1}{M_{\rm B_s} \Gamma_{\rm B_s}}\, \langle \overline{B}_s |
    {\rm Im}\, {\cal T} | B_s \rangle \ \ , \ \ \mbox{where}
\\
{\cal T} \ &=& \
    \,i\! \int\! {\rm d}^4 x\, T \left(
    {\cal H}^{\Delta B=1}_{SM} (x)\, {\cal H}^{\Delta B=1}_{NP}(0)
\right) \ \ .\nonumber
\end{eqnarray}
where we represent the generic NP $\Delta B=1$ Hamiltonian ${\cal H}^{\Delta B=1}_{NP}$ as
\bea\label{HamNP} && {\cal H}^{\Delta B=1}_{NP} = \sum_{q,q'} \
D_{qq'} \left[\overline {\cal C}_1(\mu) Q_1 + \overline {\cal C}_2
(\mu) Q_2 \right]\ ,
\\
&& Q_1 =  \overline{b}_i \overline\Gamma_1 q_i' ~
\overline{q}_j \overline\Gamma_2 s_j\ ,
\ \
Q_2 = \overline{b}_i \overline\Gamma_1 q_j' ~
\overline{q}_j \overline\Gamma_2 s_i\ ,
\nonumber
\eea
where the spin matrices $\overline\Gamma_{1,2}$ can have an arbitrary
Dirac structure, $D_{qq'}$ are some New Physics-generated coefficient 
functions~\cite{Golowich:2006gq}, and $\overline {\cal C}_{1,2}(\mu)$ are 
Wilson coefficients evaluated at the energy scale $\mu$. This gives us the
following contribution to the lifetime difference:
\begin{equation}
\label{New Physics result} 
\DG^{NP} = -\frac{8G_F
\sqrt{2}}{M_{B_s}}\sum_{qq'}D_{qq'}V^{\ast}_{qb}V_{q's}
\left(K_1\delta_{i j}\delta_{k l}+
K_2\delta_{k j}\delta_{i l}\right)\sum_{m=1}^{5}I_j(x,x')
\langle\overline{B_s}|O_m^{i j k l}|B_s\rangle,
\end{equation}
where $i, j, k, l$ are the color indices, $\{K_\alpha\}$ are combinations 
of Wilson coefficients,
\begin{eqnarray}
K_1= \left({\cal C}_2 \overline{\cal C}_2 N_c
+ \left({\cal C}_2 \overline{\cal C}_1 + \overline{\cal C}_2
{\cal C}_1 \right)\right), \
K_2 = {\cal C}_1 \overline{\cal C}_1 \ \
\end{eqnarray}
with the number of colors $N_c = 3$, and the operators
$O_m^{i j k l}$ are the following:
\begin{eqnarray}
\label{New Physics operators}
O_1^{i j k l} &=& \left(\bar{b}_i\Gamma^{\nu}\gamma^{\rho}
\Gamma_2s_{l}\right)\left(\bar{b}_k\Gamma_{1}
\gamma_{\rho}\Gamma_{\nu}s_{j}\right)
\nonumber \\
O_2^{i j k l} &=& \left(\bar{b}_i\Gamma^{\nu}\not{p}\Gamma_2s_{l}\right)
\left(\bar{b}_k\Gamma_{1}\not{p}\Gamma_{\nu}s_{j}\right)
\nonumber \\
O_3^{i j k l} &=& \left(\bar{b}_i\Gamma^{\nu}\Gamma_2s_{l}\right)
\left(\bar{b}_k\Gamma_{1}\not{p}\Gamma_{\nu}s_{j}\right),
\\
O_4^{i j
k l} &=& \left(\bar{b}_i\Gamma^{\nu}\not{p}\Gamma_2s_{l}
\right)\left(\bar{b}_k\Gamma_{1}\Gamma_{\nu}s_{j}\right)
\nonumber \\
O_5^{i j
k l} &=& \left(\bar{b}_i\Gamma^{\nu}\Gamma_2s_{l}\right)
\left(\bar{b}_k\Gamma_{1}\Gamma_{\nu}s_{j}\right)\nonumber,
\end{eqnarray}
where $\not{p}$ is the $b$-quark momentum operator.
Defining $z_q\equiv m_q^2/m_b^2$ and $z_{q'}\equiv m_{q'}^2/m_b^2$
the coefficients $I_j(z_q,z_{q'})$ can be written as follows:
\begin{eqnarray}
I_1(z_q,z_{q'}) &=& -\frac{k^* m_b}{48\pi}\left[1-2(z_q+z_{q'})+(z_q-z_{q'})^2\right]
\nonumber,\\
I_2(z_q,z_{q'}) &=& -\frac{k^*}{24m_b\pi}\left[1+(z_q+z_{q'})-2(z_q-z_{q'})^2\right]
\nonumber,\\
I_3(z_q,z_{q'}) &=& \frac{k^*}{8\pi}\sqrt{z_q}\left[1+z_{q'}-z_q\right],
\\
I_4(z_q,z_{q'}) &=& -\frac{k^*}{8\pi}\sqrt{z_{q'}}\left[1-z_{q'}+z_q\right]
\nonumber,\\
I_5(z_q,z_{q'}) &=& \frac{k^* m_b}{4\pi}\sqrt{z_q z_{q'}}\nonumber,
\end{eqnarray}
where
$k^*=(m_b/2)\left[1-2(z_q+z_{q'})+(z_q-z_{q'})^2\right]^{1/2}$. This is the 
most general formula for the New Physics contribution to the lifetime difference in 
$B_s$ mesons. We now look into two particular examples extensions of the Standard Model,
multi-doublet Higgs models and Left-Right Symmetric Models, that can contribute to $\DG$.

\subsection{Multi-Higgs model}\label{Multi-Higgs}

One of possible realizations of New Physics is a multi-Higgs doublet 
model~\cite{Glashow:1976nt}. Many of SM extensions, particularly the 
supersymmetric ones, require extended Higgs sector in order to 
break additional symmetries of NP down to $SU(2)_L \times U(1)$ of the 
Standard Model. These constructions contain charged Higgs bosons as parts 
of the extended Higgs sector. These models provide new flavor-changing 
interactions mediated by charged Higgs bosons, which lead to rich
low-energy phenomenology~\cite{Barger:1989fj,Atwood:1996vj}. In the 
low-energy limit, charged Higgs exchange leads to the following four-fermion 
interaction~\cite{Golowich:1979hd},
\begin{equation}
\label{Higgs doublet hamiltonian} {\cal{H}}^{\Delta
B=1}_{ChH}=-\frac{\sqrt{2}G_F}{M_H^2}\ \overline{b}_i
\overline\Gamma_1 q_i' ~ \overline{q}_j \overline\Gamma_2 s_j\, ,
\end{equation}
where $\overline{\Gamma}_i, \ i=1,2,$ are
\begin{eqnarray}
\overline\Gamma_1 &=& m_bV_{cb}^{\ast}\cot\beta P_L - m_cV_{cb}^{\ast}\tan\beta P_R\nonumber,\\
\overline\Gamma_2 &=& m_sV_{cs}\cot\beta P_R - m_cV_{cs}\tan\beta
P_L,
\end{eqnarray}
where $P_{L,R}=(1\mp\gamma_2)/2$. Inserting Eq.(\ref{Higgs doublet hamiltonian}) 
into Eq.~(\ref{New Physics result})
leads to a contribution to the lifetime difference $\left(\DG/\Gamma_s\right)_{ChH}$ from
three operators with various coefficients, 
\begin{eqnarray}
\left. \frac{\DG}{\Gamma_{B_s}}\right|_{ChH} = \frac{16 G_F^2m_b^2}{M_B\Gamma_{B_s}}
\frac{(V_{cb}^{\ast}V_{cs})^2}{M_H^2} \ 
& & \left[ \langle Q_1\rangle \left(4 K_2 \sqrt{z_s} I_1 \cot^2\beta + 
2 (\cot^2\beta m_b^2 \sqrt{z_s} I_2 - m_b \sqrt{z_c} I_4) (K_2 - K_1) \right) \right.
\nonumber\\
&+& \left.\langle Q_2\rangle\left(-2 K_1 \sqrt{z_s} I_1 \cot^2\beta +
(\cot^2\beta m_b^2 \sqrt{z_s} I_2 - m_b \sqrt{z_c} I_4) (K_2 - K_1)\right) \right.
\nonumber\\
&+& \left.\langle Q_3\rangle (K_1 + K_2) \left(z_c \tan^2\beta
I_5 - m_b \sqrt{z_c} I_3 \right)\right].
\end{eqnarray}
$I_i\equiv I_i(z_c,z_c)$, $K_i$ are defined above, and $\langle Q_i\rangle$ are 
\begin{eqnarray}
Q_1 &=& \left({\overline b_i}_L {s_i}_R\right)
\left({\overline b_k}_R {s_k}_L\right), \qquad 
\langle Q_1\rangle =-\frac{1}{4}f_B^2M_B^2\frac{M_B^2}{(m_b+m_s)^2}
\left(2+\frac{1}{N_c}\right)
\nonumber \\
Q_2 &=& \left({\overline b_i}_R \gamma^{\nu} {s_i}_R\right) 
\left({\overline b_k}_L \gamma_{\nu} {s_k}_L\right), \ \ \ 
\langle Q_2\rangle =
-\frac{1}{2}f_B^2M_B^2\left(1+\frac{2}{N_c}\right),
\\
Q_3 &=& \left({\overline b_i}_L \gamma^{\nu} {s_i}_L \right) 
\left({\overline b_k}_L\gamma_{\nu} {s_k}_L \right), \ \ \ 
\langle Q_3\rangle =\frac{1}{2}f_B^2M_B^2\left(1+\frac{1}{N_c}\right).
\nonumber
\end{eqnarray}
For values of $M_H = 85\,GeV$ and $\cot\beta=0.05$ \cite{PDG} we
obtain $\left(\DG/\Gamma_s\right)_{ChH} \approx 0.006$. This is about 6\% of
the Standard Model value, too small to constrain the model from 
this observable. The dependence of $\left(\DG/\Gamma_s\right)_{ChH}$ on
the mass of the Higgs boson is given in Fig.~\ref{fig:Higgs}.
\begin{figure}
\centerline{\psfig{file=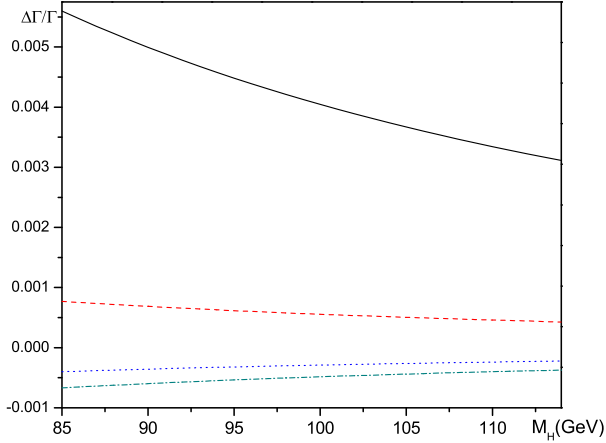,width=9cm}}
\centerline{\parbox{10cm}{\caption{\label{fig:Higgs} Dependence of
$y_{ChH}$ on the mass of the Higgs boson. Solid line:
$\tan\beta=20$; dashed line: $\tan\beta=10$; dotted line:
$\tan\beta=5$; dash-dotted line: $\tan\beta=3$.}}}
\end{figure}
%

\subsection{Left-Right Symmetric Models}

One of the puzzling features of the Standard Model is the left-handed 
structure of the electroweak interactions. A possible extension of the SM, a 
Left-Right Symmetric Model (LRSM) assumes the extended $SU(2)_L\times SU(2)_R$ 
symmetry of the theory, which restores parity at high 
energies~\cite{Mohapatra:1974gc}. While in the simplest realizations of LRSM the 
right-handed symmetry is broken at a very high scale, models can be 
consistently modified to yield $W_R$-bosons whose masses are not far above 
1 TeV range~\cite{Kiers:2005gh}. In this case flavor-changing interaction from $W_R$-bosons 
can affect $\DG$ (for a similar effect in $D$-mixing, see~\cite{Golowich:2006gq}).

In principle, manifest left-right symmetry requires that couplings to 
left-handed particles to be the same as the once to the right-handed 
ones, {\it e.g.} $g_L=g_R$. This also assumes that the right-handed 
CKM matrix $V_{ik}^{(R)}$ should be the same as the left-handed CKM matrix
$V_{ik}$. In this case, kaon mixing constraints exclude 
$M_{W_R} < 1.6$~TeV~\cite{Beall:1981ze} (direct constraints are weaker by 
approximately factor if two). However, $V_{ik}^{(R)}$ could also be quite different from 
the $V_{ik}$, as long as it is still unitary. 
In this case of non-manifest left-right symmetry the bounds on $M_{W_R}$ are
significantly weaker, $M_{W_R} > 0.3$~TeV from kaon mixing~\cite{Olness:1984xb}. 
To assess the contribution from $W_R$ to $\DG$, we equate
\begin{eqnarray}
D_{qq'} = V_{cb}^{\ast(R)}V_{cs}^{(R)}\frac{G_F^{(R)}}{\sqrt 2},
\qquad \overline \Gamma_{1,2} = \gamma^{\mu}P_R
\end{eqnarray}
in Eq.~(\ref{New Physics result}) and evaluate the respective operators. 
Here ${G_F^{(R)}}/{\sqrt 2}=g_R^2/8M^{(R) 2}_W$, and we assume $g_R = \kappa g_L$. 
In the studies of non-manifest LRSM, we shall also assume 
$\kappa = 1, 1.5, 2$~\cite{Chang:1984uy}. At the end, LRSM gives the following 
contribution to the value of ${\DG/\Gamma_{B_s}}$:
\begin{equation}
\left. \frac{\DG}{\Gamma_{B_s}}\right|_{LR} =
-V_{cb}^{\ast}V_{cs}V_{cb}^{\ast(R)}V_{cs}^{(R)}\
\frac{2 \kappa^2 G_F^2m_b^2 z_c \sqrt{1-4 z_c}}{\pi
M_B \Gamma_{B_s}}\left(\frac{M_W}{M_W^{(R)}}\right)^2\left[C_1 \langle
Q_2\rangle - 2 C_2 \langle {Q_1}\rangle\right].
\end{equation}
The dependence of $(\DG/\Gamma_{B_s})_{LR}$ on the mass of the $W_R$ boson is given in 
Fig.~\ref{fig:LR}. We see that contrary to the $D$-meson 
case~\cite{Golowich:2006gq,Golowich:2007ka}, $B_s$-mixing could provide decent 
constraints on the values of $M_W^{(R)}$. For instance, in a 
non-manifest LRSM (with relevant $V_{ij}^{(R)}\approx 1$), $\kappa=1$, and 
$M_W^{(R)}=1\ TeV$, one obtains $(\DG/\Gamma_{B_s})_{LR} \simeq -0.04$ 
This is a rather large contribution to $\DG$, more than a third of the absolute value
of the Standard Model contribution and of the opposite sign. The LRSM contributions for
$\kappa > 1$ are even larger. As expected, in the case of manifest 
LRSM ($V_{ij}^{(R)}= V_{ij}$) the contribution from this model is less marked, 
$(\DG/\Gamma_{B_s})_{LR} < 0.002$ for $M_W^{(R)} > 800$~GeV. 
\begin{figure}
\centering
\subfigure[Dependence of $\DG/\Gamma_s$  on $M_W^{(R)}$. 
Solid line: non-manifest LRSM ($\kappa=1$), dashed line: manifest LRSM.] 
{
    \label{fig:LRa}
    \includegraphics[width=7.5cm]{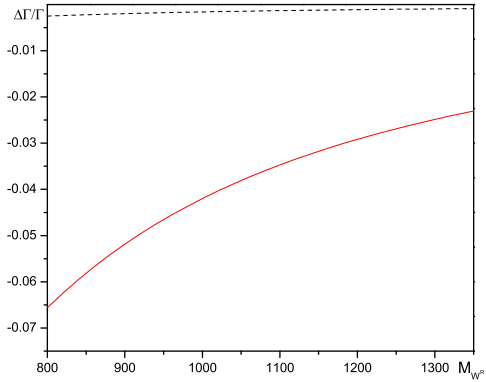}
}
\hspace{1cm}
\subfigure[Dependence of $\DG/\Gamma_s$ on $M_W^{(R)}$ in 
non-manifest LRSM. Solid line: $\kappa=1$, dashed line: $\kappa=1.5$, 
dotted line: $\kappa=2$.] 
{
    \label{fig:LRb}
    \includegraphics[width=7.5cm]{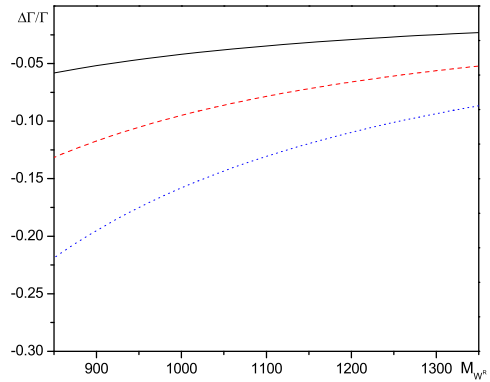}
}
\caption{Contributions to $\DG/\Gamma_s$ in the Left-Right Symmetric Models.}
\label{fig:LR} 
\end{figure}

\section{Conclusions}\label{Conclusions}

We computed the subleading $1/m_b^2$ corrections to the
difference in the lifetimes of $B_s$ mesons. We showed that they can be parameterized 
by 13 nonperturbative parameters, which we denote $\alpha_i$ and $\beta_i$.
We adopted the statistical approach for presenting our results and generate 
100000-point probability distributions of the lifetime difference, 
obtained by randomly varying our parameters within a $\pm 30\%$ 
interval around their ``factorization'' values, except for the case when 
the parameters are known from lattice QCD. In this case they are taken 
to vary within a $1\sigma$ interval as indicated above.

The results are presented in Fig.~(\ref{fig:SM}). While there is no theoretically-consistent 
way to translate the histogram of Fig.~\ref{fig:SM} into numerical predictions 
for $\DG/\Gamma_s$, we provide an estimate by taking the width of the distribution Fig.~\ref{fig:SM} 
at the middle of its height as 1-$\sigma$ variance and position of the maximum of the 
curve as the most probable value, 
\beq
\DG=0.072^{+0.034}_{-0.030} ~\mbox{ps}^{-1}, \quad \frac{\DG}{\Gamma_{B_s}} = 0.104 \pm 0.049,
\eeq
The effects of $1/m_b^2$ corrections to calculations of $\DG$ are shown to be small.

We also looked into $\Delta B = 1$ New Physics contribution to the lifetime difference 
in the $B_s$ system. We have shown that these contributions can both enhance or reduce 
the Standard Model contribution. We considered the most general four-fermion effective 
Hamiltonian, which can be generated by any reasonable extension of the Standard Model and
derived its contribution to $\DG$. We then evaluated effects of charged Higgses and  
right-handed W's on the lifetime difference. While the contribution of charged Higgs
was shown to be negligible in $\DG$, LRSM can be constrained with measurement of
$\DG$, provided lattice or QCD sum rule community provide better estimates of non-perturbative
parameters entering the SM calculation of the lifetime difference in $B_s$ mesons.

\section*{Acknowledgments}

This work was supported in part by the U.S.\ National Science Foundation 
CAREER Award PHY--0547794, and by 
the U.S.\ Department of Energy under Contract DE-FG02-96ER41005.


\end{document}